\pdfoutput=1

\documentclass[10pt,a4paper,twocolumn, twoside]{article}
\usepackage[pdftex]{graphicx}

\usepackage[T1]{fontenc}
\usepackage[latin1]{inputenc}

\usepackage{amsmath}
\usepackage{caption}

\usepackage{booktabs}

\makeatletter
\renewcommand{\section}{\@startsection{section}{2}{0cm}{-\baselineskip}
{0,5\baselineskip}{\normalsize\bfseries}}
\renewcommand{\subsection}{\@startsection{subsection}{3}{0cm}{-\baselineskip}
{0,5\baselineskip}{\normalsize\slshape}}
\makeatother

\newcommand{\Xe}{$^{133}$Xe}
\newcommand{\Xem}{$^{131m}$Xe}
\newcommand{\Rn}{$^{222}$Rn}
\newcommand{\Kr}{$^{85}$Kr}

\begin{document}

\title{Detection of \Xe\ from the Fukushima nuclear power plant in the upper troposphere above Germany}

\author{Hardy Simgen$\rm ^a$, Frank Arnold$\rm ^{a, b}$, Heinfried Aufmhoff$\rm ^b$, Robert Baumann$\rm ^b$, \\ Florian Kaether$\rm ^a$, Sebastian Lindemann$\rm ^a$, 
Ludwig Rauch$\rm ^a$, \\ Hans Schlager$\rm ^b$, Clemens Schlosser$\rm ^c$, Ulrich Schumann$\rm ^b$}

\date{\small \it 
$^a$Max-Planck-Institut f\"ur Kernphysik, Saupfercheckweg 1, D-69117 Heidelberg, Germany \\
$^b$DLR Oberpfaffenhofen, M\"unchner Stra{\ss}e 20, D-82234 We{\ss}ling, Germany \\
$^c$Bundesamt f\"ur Strahlenschutz, Rosastra{\ss}e 9, D-79098 Freiburg, Germany \\
\vspace{0.3cm}
Email-addresses: \\ {\tt Hardy.Simgen@mpi-hd.mpg.de \\
 Hans.Schlager@dlr.de \\ cschlosser@bfs.de} \\
\vspace{0.3cm}
{\it (Published in Journal of Environmental Radioactivity, Volume 132
  (June 2014) Pages 94-99; doi:10.1016/j.jenvrad.2014.02.002)}}

\twocolumn[
\begin{@twocolumnfalse}
\maketitle

\begin{abstract}
\noindent After the accident in the Japanese Fukushima Dai-ichi nuclear power
plant in March 2011 large amounts of radioactivity were released and
distributed in the atmosphere. Among them were also radioactive noble gas isotopes which can be used as tracers to test global atmospheric circulation models. This work presents unique measurements
of the radionuclide \Xe\ from Fukushima in the upper troposphere above Germany.
The measurements involve air sampling in a research jet aircraft
followed by chromatographic xenon extraction and ultra-low background gas
counting with miniaturized proportional counters.
With this technique a detection limit of the order of 100 \Xe\ atoms in
litre-scale air samples (corresponding to about 100 mBq/m$^3$) is
achievable. Our results provide proof that the \Xe-rich ground level air layer from
Fukushima was lifted up to the tropopause and distributed hemispherically.
Moreover, comparisons with ground level air measurements indicate that the
arrival of the radioactive plume at high altitude over Germany occurred several
days before the ground level plume.


\end{abstract}
\end{@twocolumnfalse}
]

\section{Introduction}

One of the strongest earthquakes in Japanese history happened close to the
east coast of Honshu Island on March 11, 2011 and was followed by a
destructive tsunami. This triggered a series of accidents in the nearby
Fukushima Dai-ichi nuclear power plant \cite{IAEA_news} which caused a
release of large amounts of radionuclides \cite{sourceterm1, sourceterm2,
sourceterm3, Stohl2012}, among them the radioactive xenon nuclide
\Xe. The total \Xe\ source term was estimated to be between 12 EBq and 19
EBq \cite{Stohl2012, Stohl-Xe, Bowyer}. Such a strong release provides an opportunity to test and improve global atmospheric circulation models using \Xe\
as a tracer.  Since xenon is a noble gas it is inert and rarely reacts
with other elements. Thus, it is superior to other commonly used tracers
which require profound understanding of atmospheric chemistry, washout
processes and rain. Moreover, its lifetime of 7.57 days \cite{nucdat} is comparable to the time scale for intercontinental transport in the
upper troposphere. In this work we present the detection of airborne \Xe\ in the upper
troposphere above Germany after the Fukushima event. A second paper \cite{DLR-in-prep} deals with the atmospheric aspects and
implications of the \Xe\ measurements as well as with the interpretation
of simultaneously measured trace gases and aerosols. For comparison with data obtained from ground level measurements in Europe of iodine and cesium radionuclides we refer to \cite{arrival}.

\begin{figure*}[t]
\begin{center}
\includegraphics[width=0.8\textwidth]{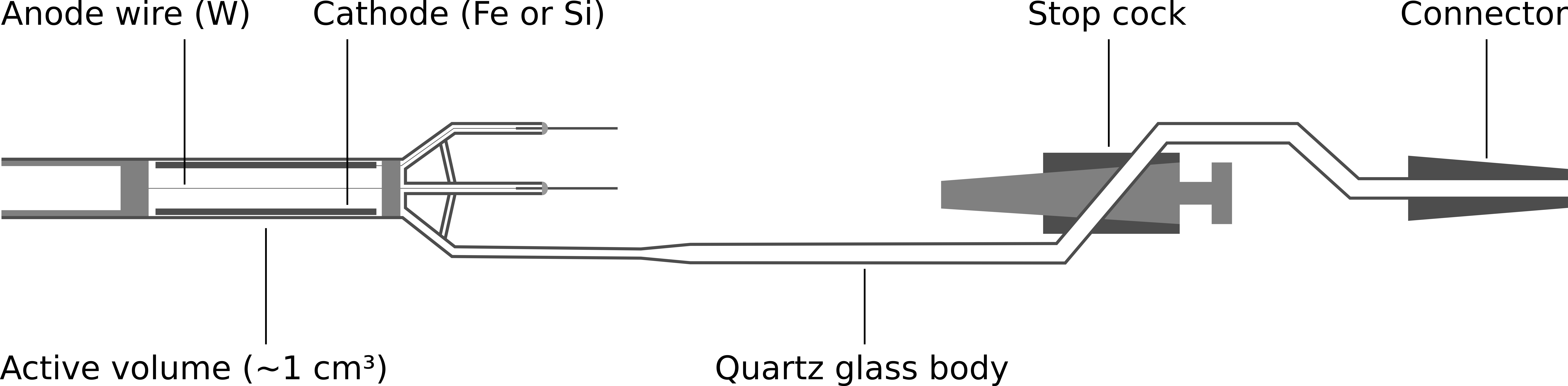}
\caption[The miniaturized proportional counter]{Sketch of the miniaturized ultra-low background proportional counter used for the \Xe\ measurements. \label{fig:counter}} 
\end{center}
\end{figure*}

After first information about the nuclear accident became public we
prepared an aircraft campaign at short notice.  The rationale of our
aircraft measurements was to investigate pollution of the upper
troposphere, particularly the tropopause region, by pollutants released
from fossil fuel combustion in the region of North-East China, Korea and Japan. In
fact, this region represents the strongest SO$_2$-source region worldwide. 
Since SO$_2$ is an important precursor of potentially climate-active
atmospheric aerosol particles, above mentioned region became a matter of
particular interest at the Deutsches Zentrum f\"ur Luft- und Raumfahrt (DLR)
\cite{DLR-in-prep, Fiedler1, Fiedler2, Schlager}.

Samples were taken during two flights with the Falcon
research jet aircraft \cite{Falcon} of DLR.
The first flight on March 23, 2011 aimed to the intercept the
Fukushima \Xe\ plume in the upper troposphere above Germany, shortly after
its expected arrival on March 21/22, 2011.  Due to strict air safety
regulations only rather small (1 litre) and passive air samplers were
approved in the short time before the first flight. The second flight on
April 14, 2011 aimed to probe the highly diluted Fukushima plume in the
upper troposphere. Here larger samples of about 10 litres could be taken. Model simulations conducted
by the Institut f\"ur Physik der Atmosph\"are of DLR prior to the
flights, predicted that upper tropospheric \Xe\ activity concentrations
above Germany would be around 1~Bq/m$^3$ during the first flight and about
10 times less during the second flight. Thus, extremely sensitive \Xe\
detection techniques in the laboratory are required to observe the
Fukushima signal in litre-scale samples. 


\section{$^{133}$Xe measurements with miniaturized proportional counters}
\label{extraction}

Miniaturized ultra-low background proportional counters (see Figure
\ref{fig:counter}) were developed at the Max-Planck-Institut f\"ur
Kernphysik  to detect few atoms of $^{71}$Ge in the {\sc Gallex} solar
neutrino experiment \cite{Wink93}. They were also used for \Xe\ studies in
{\sc Gallex} \cite{pezzoni} and later to detect low
levels of various radioactive noble gas isotopes in the astroparticle physics
experiments {\sc Borexino} \cite{borex}, {\sc Gerda} \cite{gerda} and {\sc Xenon} \cite{xenon}. Due to their low background and low energy threshold they are
ideally suited to detect gaseous radioactive isotopes with high sensitivity. The key technology is a dedicated processing of gas samples
before loading them into the counter. It includes removal of trace
impurities which disturb the performance of the proportional counter (e.g.
oxygen, humidity, ...) as well as separation of unwanted radioactive
impurities. Finally, the size of any gas sample has to be reduced without
losing the target isotope and without introducing contamination to fit into
the active volume of the miniaturized counter ($\sim1$~cm$^3$).  The sample preparation is done by means
of a dedicated gas handling and counter filling line made from glass. Such a
system was developed in the framework of astroparticle physics experiments
and as indicated in \cite{simgen_Wien} it may be used for xenon processing.

The main challenge is the xenon separation at high acceptance from the air
samples. This is done by pumping the sample over a 0.6 gram activated carbon
column held at -186 $^\circ$C by immersing it in liquid argon.  While most
of the oxygen and nitrogen is pumped away, xenon is efficiently stopped in
the column.  Subsequently, the sample is transferred to the top of a gas
chromatography column by heating the activated carbon column and cooling
the gas chromatography column at liquid nitrogen temperature. Helium is
used as a carrier gas to run the sample through the gas chromatography
column which is filled with Chromosorb 102 \cite{Chromosorb}. 
Nitrogen, oxygen, carbon dioxide, xenon and radon elute from such a column
in the listed order.
The temperature is varied step-wise from liquid nitrogen
temperature to -110 $^\circ$C and finally to -30 $^\circ$C to achieve good
separation of all components.   
A big challenge is \Rn\ in the sample eluting shortly after xenon. Radon is
not visible in the chromatogram which is recorded by a thermal
conductivity detector with limited sensitivity.
Thus, it is necessary to reject the late fraction of the xenon which might not be
fully separated from \Rn.  This causes losses of \Xe\ between five and ten percent
for a single run. These losses are quantified via a natural xenon carrier which is
added to the sample beforehand.  Xenon has no long-lived
radioactive isotopes, therefore the carrier which is several years old
does not introduce background radioactivity. Together with 10~\% of the
quenching gas methane the xenon is used as a counting gas for the
proportional counters which are operated at atmospheric pressure.  The
radon issue is discussed in more detail in section \ref{radon}.\\

The energy calibration is done by illuminating the proportional counters
with cerium X-rays according to a technique which was developed for the {\sc
Gallex} experiment \cite{diss_urban}. The cerium X-rays excite xenon atoms
and generate a discrete low-energy line spectrum in the proportional
counter. The energy-to-channel allocation is done by fitting a second order
polynomial to the positions of the three peaks that appear at 1.1 keV, 5.0 keV and 9.7 keV.
For some runs the full absorption peak centered at 34.6 keV was included in the fit and it was confirmed
that a second order polynomial provides a good description up to that
energy. Another peak at 0.3 keV was not considered, because it was not
always clearly resolved due to its vicinity to the threshold.


\section{Counting system}
\label{counting_sys}

\begin{figure}
\begin{center}
\includegraphics[width=\columnwidth]{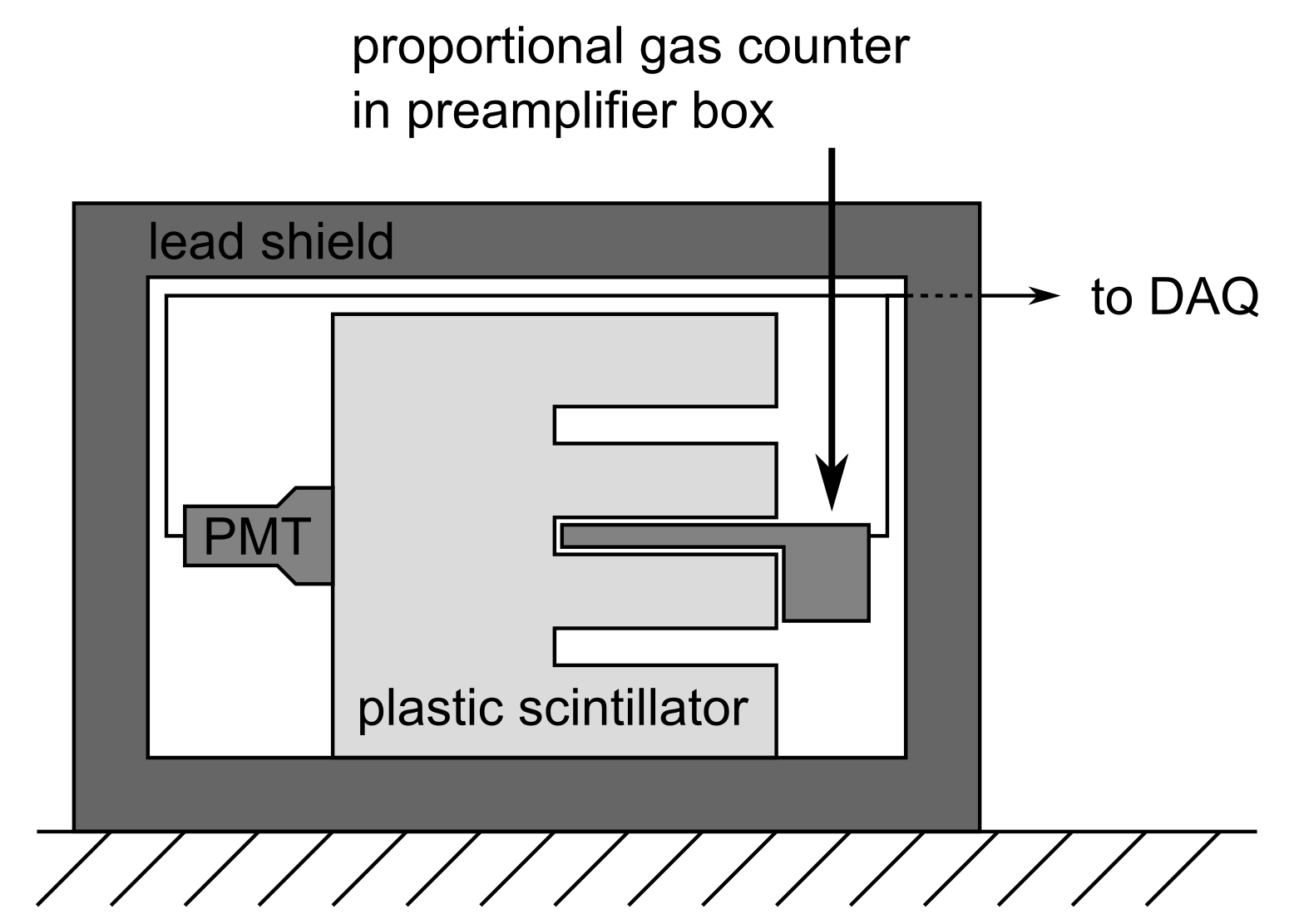}
\caption[Sketch of the gas counting setup]{Sketch of the gas counting setup with passive shield and active plastic scintillator veto system. \label{fig:counting} } 
\end{center}
\end{figure}

The counting system (see Figure \ref{fig:counting}) to read out the
proportional counters is located at the subterranean Low-Level-Laboratory of the
Max-Planck-Institut f\"ur Kernphysik in Heidelberg, which is at a depth
corresponding to about 15 m of water equivalent.  The counters are placed
in a plastic scintillator veto system which provides almost 4-$\pi$ coverage
to reject cosmic muons.  The plastic scintillator itself is embedded in a
15 cm lead shield.

\Xe\ decays with a lifetime of 7.57 days by beta disintegration with a
Q-value of 427.4 keV (see Table \ref{tab:decaydata}).  99.12~\% of all decays populate a
81~keV isomeric state (lifetime 9.1 ns) which de-excites with 37~\%
probability by emission of a gamma quant and in the remaining cases by
internal conversion \cite{nucdat}.  In the proportional counters we measure the energy
deposition of emitted electrons and of X-rays from simultaneous
de-excitation of the atomic shell.  Typically, only a small fraction of
the released energy is deposited in the miniaturized counters, thus a low
energy threshold is required.  At around 150 eV the efficiency of our data acquisition 
system drops down significantly.  To avoid threshold effects a conservative 500 eV
cut-off was applied in the data analysis.  The 81~keV gamma ray might trigger
an unwanted veto signal in the plastic scintillator.  Thus, the counters
were directly surrounded by a $\sim5$ mm lead shield to which a box
containing the frontend electronics is connected. Signals  are read out by
a Flash - Analog to Digital (FADC) converter board (Struck SIS3301) at 100
MHz sampling rate and 14 bit resolution.  The rise-time of a proportional
counter pulse is defined by the time difference between 10~\% and 90~\% of
its maximum amplitude.  The
plastic scintillator covers a surface of approximately 0.6 $\rm{m^2}$
perpendicular to the muon flux resulting in a muon event rate of about
100~Hz.  Each signal from the proportional counter that is not in
coincidence with a plastic scintillator signal is recorded with the
indication of timestamp, energy and rise-time.  
 \Xe-events have a rise-time of $\sim 1~\mu$s.  A
conservative rise-time cut of $3~\mu$s was applied to discriminate against
a population of slow pulses which was present in some of the data.

\begin{table*}[t]
\centering
\begin{tabular}{cccc}
  \toprule
 
   Isotope & Lifetime $\tau$ & Q-value [keV] & Decay-mode\\

  \midrule
        \Xe  & 7.57 d & 427.4 & $\beta$-decay to excited levels\\
        \Xem & 17.21 d & 163.9 & Converted gamma transition\\
        \Kr & 15.51 a & 687.1 & $\beta$-decay ($0.43\%$ to excited level)\\
        \Rn & 5.52 d & 5590.3 & $\alpha$-decay\\
 
\end{tabular}
\caption[Decay data]{Selected decay data of relevant radionuclides taken from \cite{nucdat}.}
\label{tab:decaydata}
\end{table*}


\section{Efficiency calibration}

\label{efficiency}
\begin{figure}
\begin{center}
\includegraphics[width=\columnwidth]{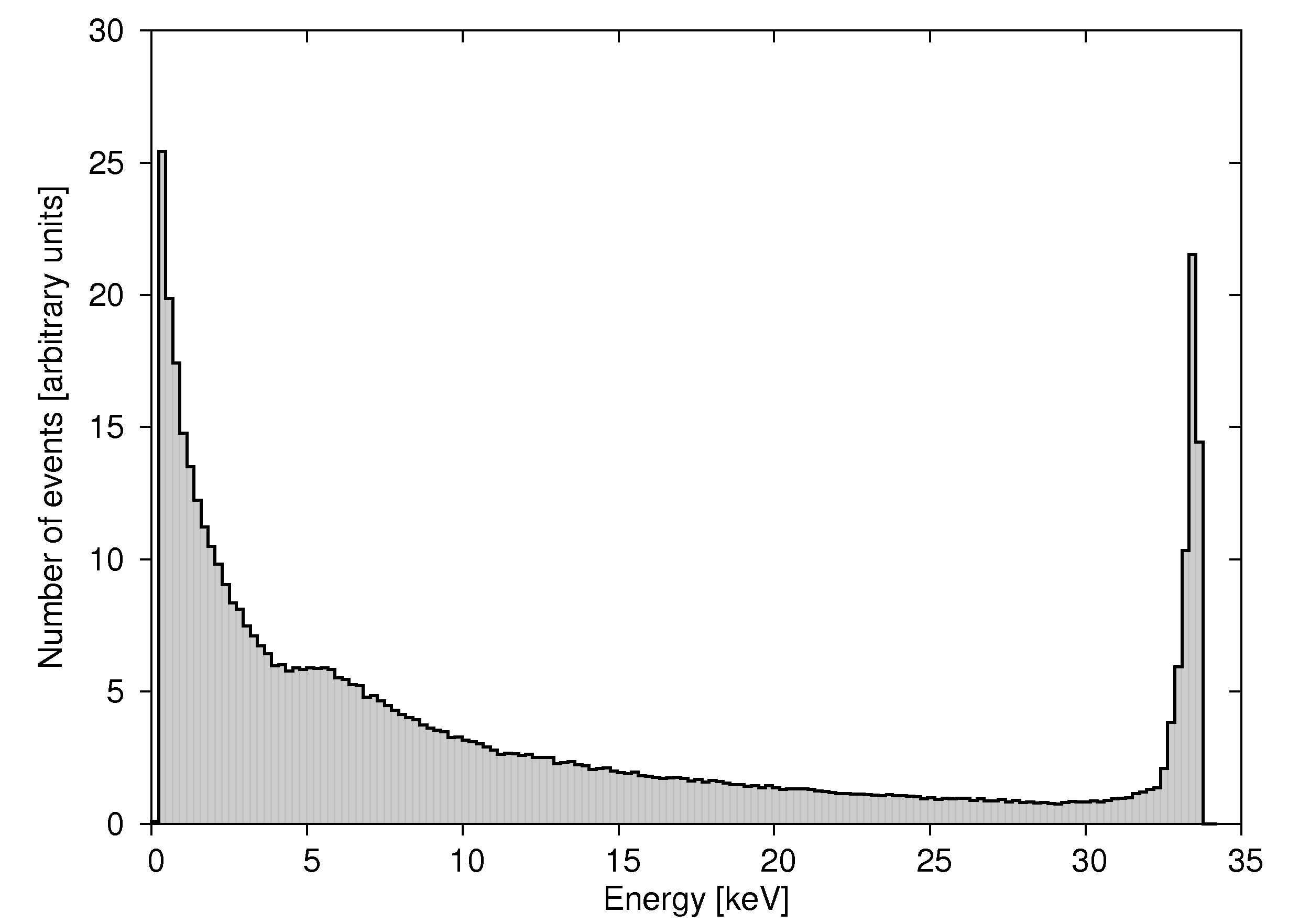}
\caption[\Xe\ efficiency calibration spectrum]{Spectrum of a \Xe\ standard for efficiency calibration. The bump around 5~keV is due to admixed \Xem, which is present at a few percent in the \Xe\ standard. \label{Fig:calibration} } 
\end{center}
\end{figure}

A commercial \Xe\ calibration standard was used to determine the detection
efficiency of the proportional counters for \Xe.  It was first measured
with a high purity germanium (HPGe) gamma spectrometer to identify unwanted
radionuclides.  Besides \Xe, the gamma rays from \Xem\ and \Kr\ were
visible in the spectrum indicating contamination with these
nuclides at the percent level. Figure \ref{Fig:calibration} shows an
energy spectrum of the calibration standard as recorded with one of the
proportional counters.

Alpha-particles deposit a large amount of energy in the counter due to the
high ionization density along their track. Thus, overflow channels are not
considered in the analysis and an upper energy threshold of 27~keV is
applied to remove these events.

Using the selection criteria discussed in section \ref{counting_sys}
good events are selected in a window between 0.5~keV and 27~keV and with a
rise-time of less than  $3~\mu$s.  To obtain the efficiency for \Xe, the
temporal decay curve of these events is fitted with two exponential decays
for \Xe\ and \Xem, respectively and a constant background component, which
comprises mainly long-lived \Kr. 
In the fit the lifetimes of \Xe\ and \Xem\ were fixed to the literature values given in Table \ref{tab:decaydata}.
The obtained efficiencies for \Xe\ vary between 49~\% and 56~\% for the six counters in use. 
Such variation is expected due to geometrical differences of the counters
resulting in different ratios of active to passive volumes.

\section{Maximum likelihood analysis}\label{data_analysis}

The recorded events of all samples were analyzed with a maximum likelihood
approach which was originally developed for data analysis in radiochemical
solar neutrino experiments \cite{Cle83}. To estimate the initial number
$N$ of $^{133}$Xe atoms in a given sample the probability density function
of $^{133}$Xe decays is used which decreases exponentially with the \Xe\ lifetime. 
In contrast, the background distribution is assumed to be constant in time
with a rate $b$.  The expected sources of background events are unvetoed
muons and contamination of the proportional counters with traces of
radioactive nuclides (e.g. $^{210}$Pb) with long lifetimes.  Contributions from
\Xem, $^{133m}$Xe and $^{135}$Xe which were also released during the
accident are negligible in the four samples of the first flight
\cite{Bowyer}. Data from the Bundesamt f\"ur
Strahlenschutz (see section \ref{results}) suggest that the \Xem\
fraction of the signal reaches 11~\% in the second flight. We checked that the inclusion of
this component in the likelihood function does not change the result
for \Xe\ significantly.  In particular, since the \Xem\ also
originates from Fukushima, it slightly improves the significance for
the detection of radioxenon from the accident. However, the efficiency
for detection of \Xem\ was not calibrated, so we decided to ignore it in the analysis.

If $t_i$ are the time stamps of the $n$ recorded events and if $T$
describes the time intervals of the total  measurement, the resulting
likelihood function $\cal L$ can be written as

\[ \log \mathcal{L}(N, b) = - \int_T \left( \frac{N}{\tau}e^{-t/\tau} +b
\right) dt + \qquad \quad \]
\begin{equation} \label{eq:log-likelihood}
\qquad \qquad \qquad +\sum _{i=1} ^n \log \left (\frac{N}{\tau}e^{-t_i/\tau}
+b\right).
\end{equation}

While the first part of this equation is derived from the
probabilities of events occurring at the time stamps $t_i$, the second part
describes the likelihood of no observed events in between.  Finally, $\log
\mathcal{L}(N, b)$ is numerically maximised to get the best estimators
$\hat N$ and $\hat b$ for the free parameters. In maximum likelihood
theory, the statistical error of a fit-parameter $a$ is estimated by
varying the parameter around its best fit value $\hat a$ until

\begin{equation} \label{eq:log-likelihood-error}
\log\mathcal{L}(\hat a) - \log \mathcal{L}(\hat a \pm \sigma_a) = \frac{1}{2}
\end{equation}

while $\log \mathcal{L}(\hat a \pm \sigma_a)$ is maximised regarding the
remaining free parameters. To consider a possible asymmetry of the error,
both sides of $\hat a$ are treated independently. However, the asymmetry
turned out to be negligible in our case.\\


\section{Radon correction}\label{rn_correction} 
\label{radon}

\begin{figure}[t]
\begin{center}
\includegraphics[width=\columnwidth]{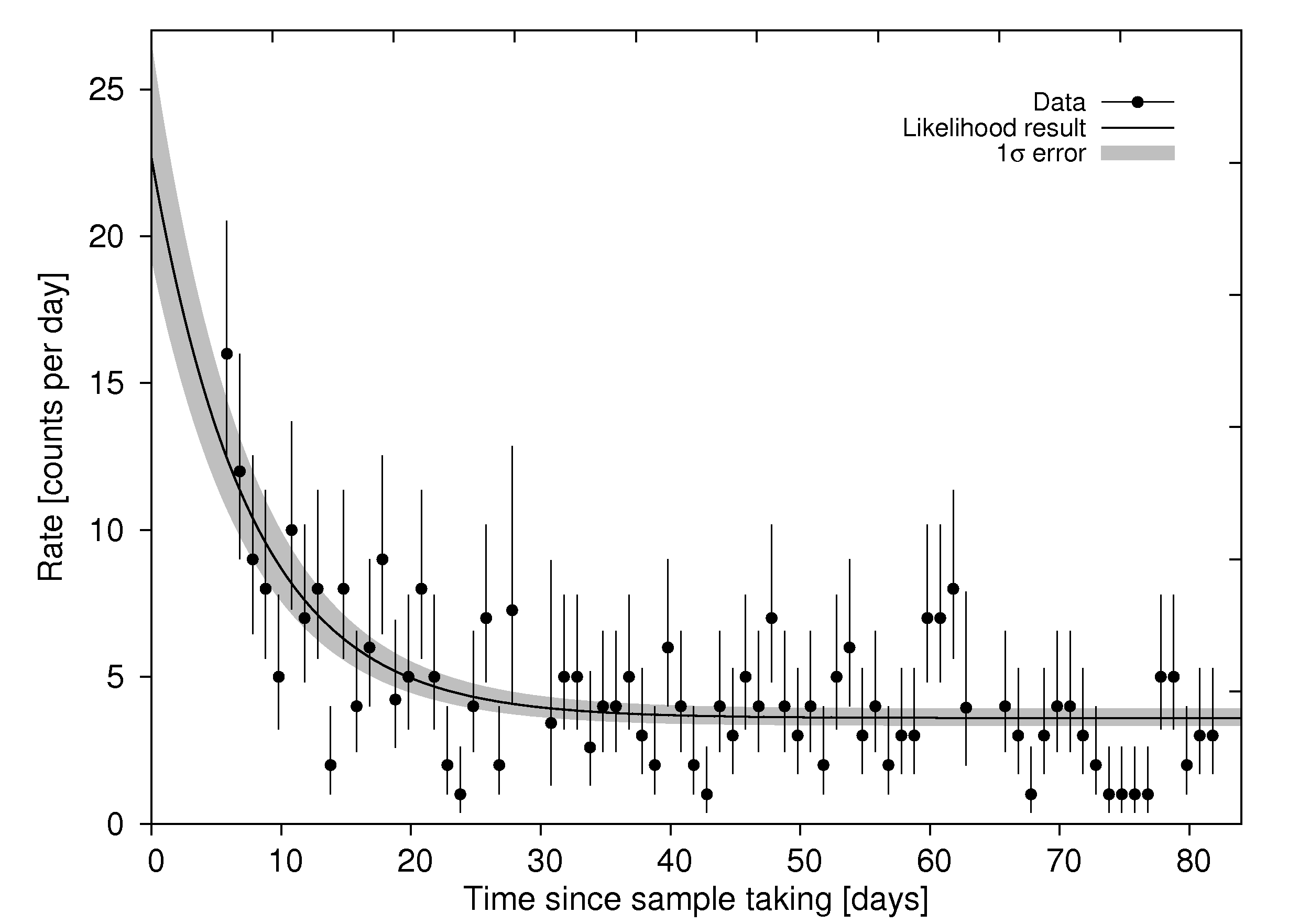}
\caption[Temporal development of decay rate]{Temporal development of the measured decay rate per day for sample 'Flight A-1'. The excess due to presence of $^{133}$Xe in
the left part of the plot is clearly visible. The black curve shows the result and the 1~$\sigma$ uncertainties (grey band) of the maximum likelihood analysis. \label{fig:DLR-2} }
\end{center}
\end{figure}

The $^{222}$Rn concentration in ambient air fluctuates depending on local
geological properties, on meteorological influences and on the altitude.
Even in the upper troposphere it might not be negligible with respect to
the expected \Xe\ signal from the Fukushima plume \cite{radon-high}. Thus,
special care was taken to separate radon from xenon during the gas
chromatography procedure (see section \ref{extraction}). However, since the
separation is difficult, it cannot be excluded that a few radon atoms
entered the proportional counter. The lifetime $\tau = 5.52 \, \rm d$
\cite{nucdat} of $^{222}$Rn is in the range of $^{133}$Xe, therefore radon
can not be treated as a part of the constant background $b$ in the
likelihood analysis. 

To investigate possible radon contamination of the samples, a delayed
coincidence analysis was performed: The $^{222}$Rn decay chain includes
$^{214}$Bi with its short-lived progeny $^{214}$Po (lifetime $237 \,
\mu{\rm s}$ \cite{nucdat}). Thus, a  $^{214}$Bi-$^{214}$Po (BiPo)
coincidence event is characterized by a $\beta$-decay of $^{214}$Bi
followed by an $\alpha$-decay of $^{214}$Po on a short time scale. With the
measurement of a $^{222}$Rn standard sample the detection of one BiPo event
was calibrated to correspond to an average of $10.6 \pm 3.7$ radon events
in the energy region of interest between 0.5 keV and 27 keV.\\

From the six measured samples five showed no BiPo event, while there were
two BiPo events in the sample 'Flight A-1' (see section \ref{results}).
Since the gas handling procedure was similar for all six samples we assume
the contamination risk to be the same for all samples. Thus, the number of
radon atoms entering the counter is expected to be Poisson distributed.
Using the quoted numbers above, a likelihood analysis of this scenario
leads to a correction of the number $N$ of $^{133}$Xe atoms in each sample
by $3.5^{+3.3}_{-2.3}$.


\section{Results}
\label{results}

In total six samples were obtained during the two flights. The first flight
took place on March 23, 2011 and four samples were collected in evacuated stainless
steel containers of one litre volume. On the second flight on April 14,
2011 we combined five samples from various places in Northern Germany and
five samples from various places in Southern Germany to obtain two
larger samples, each of about 10 litres air. All samples were taken
inside the aircraft's cabin where a constant pressure of 800 mbar is
maintained. The cabin air is permanently exchanged with outside air
and a complete exchange takes only a few minutes.

%
The samples were counted in our gas counting setup differently long,
but at least several months, such that the constant background rate
could be precisely determined after the decay of \Xe. 
As an example Figure \ref{fig:DLR-2} shows the data from sample 'Flight
A-1'. The result of the maximum likelihood analysis is given in Table
\ref{tab:fit}. For each sample the number of initial atoms at the start of
the measurement $\hat N$ and the constant background rate $\hat b$ as
described in section \ref{data_analysis} are given. Sample 'Flight B-2'
suffered from an unexpected high background which cannot be explained
easily.  Five of the six samples show a signal above the decision
threshold. For sample 'Flight A-3' an upper limit at 90\% confidence level
is given.

\begin{table}[t]
\centering
\begin{tabular}{cr@{ $\pm$ }lr@{ $\pm$ }l}
  \toprule
 
   Sample & \multicolumn{2}{c}{$\hat N$ [atoms]} & \multicolumn{2}{c}{$\hat b$ [cpd]}\\

  \midrule
        Flight A-1  & 72 & 13  &  3.5 & 0.3  \\
        Flight A-2  & 102 & 18 & 8.8 & 0.4  \\
        Flight A-3  &  \multicolumn{2}{c}{$<37$} & 5.6 & 0.3\\
        Flight A-4  & 27 & 12 & 4.1 & 0.4  \\
        Flight B-1  & 46 & 16 & 5.1 & 0.6  \\
        Flight B-2  & \hspace{1ex} 61 & 32 & 30.9 & 1.4
 
\end{tabular}
\caption[Fit results]{Results of the maximum likelihood analysis for the six samples. The errors are 1-$\sigma$ statistical uncertainties and the upper limit is given at 90~\% confidence level.}
\label{tab:fit}
\end{table}

\begin{table*}[t]
\centering
\begin{tabular}{ccccr@{ $\pm$ }l}
  \toprule
  Sample & Date & Coordinates & Altitude &  \multicolumn{2}{c}{\Xe\ activity}\\
   & & & [km] &  \multicolumn{2}{c}{[mBq/m$^3$ (STP)]}\\
  \midrule

       Flight A-1  & 23.3.11 & 48$^\circ$N, 11$^\circ$E &  8.1 & \hspace{4ex}540 & 110\\
       Flight A-2  & 23.3.11 & 48$^\circ$N, 11$^\circ$E &  9.2 & 950 & 180\\
       Flight A-3  & 23.3.11 & 50$^\circ$N, 11$^\circ$E &  9.2 &\multicolumn{2}{c}{$< 310$}\\
       Flight A-4  & 23.3.11 & 51$^\circ$N, 11$^\circ$E & 11.8 & 210 & 110\\
       Flight B-1  & 14.4.11 & Northern Germany &  8.5 & 42 & 16\\
       Flight B-2  & 14.4.11 & Southern Germany &  8.5 & 63 & 36\\

         \bottomrule
\end{tabular}
\caption[Measured \Xe\ activity concentrations]{Measured \Xe\ activity concentrations and coordinates / altitude of the sampling position.}
\label{tab:results}
\end{table*}

\begin{figure*}[t]
\begin{center}
\includegraphics[width=0.8\textwidth]{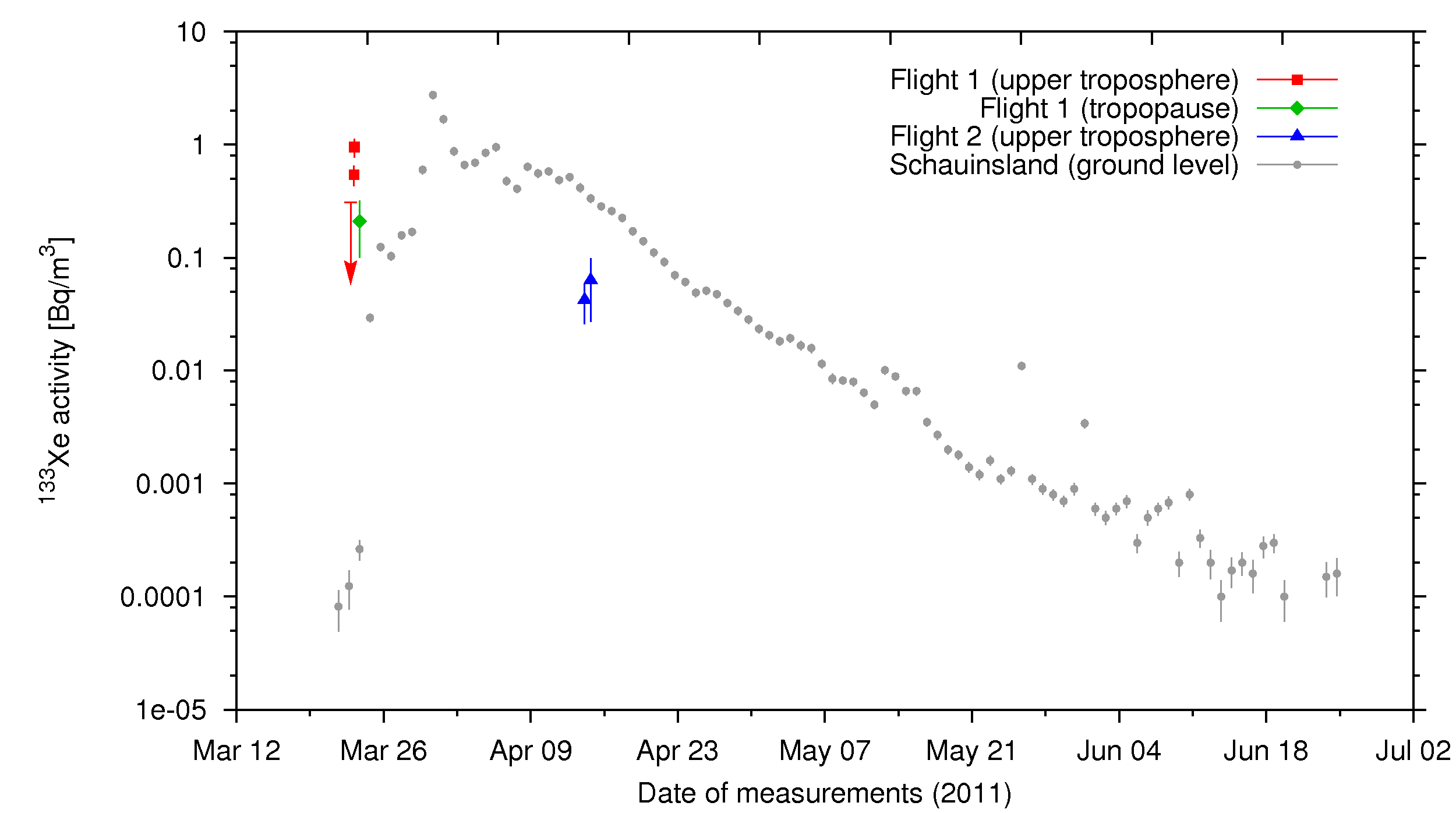}
\caption[]{Comparison of the measured \Xe\ concentration during the flights and in ground level air recorded by the Bundesamt f\"ur Strahlenschutz on the Schauinsland mountain in Southern Germany. \label{fig:Xe_BFS} }
\end{center}
\end{figure*}

The fit results are corrected for the \Rn\ contamination and then converted into a \Xe\ activity at the time of sampling by using the measured \Xe\ detection efficiency of the six proportional counters.  Also the losses as determined with the xenon carrier (see section \ref{extraction}) are taken into account. Finally, the corresponding \Xe\ activity concentration is calculated normalized to air at standard temperature and pressure (STP: 273.15~K and $10^5$~Pa).  Again the asymmetry of the 1~$\sigma$ uncertainties is ignored since it turned out to be negligible. The final results are shown in Table \ref{tab:results}.

The \Xe\ concentration in ambient air is usually well below
50~mBq/m$^3$ with rare short-term exceptions \cite{BfS}.
Consequently, we can conclude that in the first flight on March, 23 2011
clear evidence of \Xe\ from the radioactive plume of Fukushima Dai-ichi
in the upper troposphere above Germany was observed.  The evidence from
the second flight about three weeks later is weaker, since \Xe\ has
decayed and is further diluted. Nonetheless, our results for the
second flight are still above
the usual ambient activity concentration indicating that we were still detecting
the emission from Fukushima.

The Bundesamt f\"ur Strahlenschutz (BfS) operates a station for
monitoring the environmental radioactivity at the Schauinsland
mountain in Southern Germany. As Radionuclide Station 33 it is part of
the International Monitoring Network (IMS) of the Comprehensive
Nuclear-Test-Ban Treaty Organization (CTBTO) \cite{ctbto,bieringer}
and a SPALAX noble gas system \cite{Spalax} is installed to monitor
continuously the activity concentration of radioactive xenon in ground
level air. It is based on fully automated sampling of large air
samples, purification, concentration of xenon on activated carbon columns and detection by
gamma ray spectroscopy with a high purity germanium detector.  With
this system 24 h measurements of four xenon radioisotopes (\Xe, $^{133m}$Xe, \Xem\ and $^{135}$Xe) in ambient air are
performed. For \Xe\ the minimum detectable activity concentration lies
around 450~$\mu$Bq/m$^3$ (STP). Note that the air sample size is typically 60 m$^3$
(STP) which is about $10^5$ times larger than the air samples taken
during the flights in our study.  In Figure \ref{fig:Xe_BFS} results from the
Schauinsland station in the period after the Fukushima accident are shown
together with results from the two flights. At ground level the first
indication of the arrival of \Xe\ from Fukushima is visible on March 24,
2011. However, \Xe\ activity concentrations at the 1 Bq/m$^3$ level are
detected for the first time on March 29, 2011.

The relatively high \Xe\ activity concentrations measured during the first
flight on March 23, 2011 in the upper troposphere show that the arrival
of the Fukushima plume at high altitudes was earlier than at ground
level. Also remarkable is the evidence for traces of \Xe\ in the tropopause
(sample Flight A-4, green diamond in Figure \ref{fig:Xe_BFS}). Both
observations indicate that air masses from the Japanese regions were
quickly lifted to high altitudes. For more information of atmospheric
implications and comparisons with ground-level measurements the reader
is referred to
\cite{DLR-in-prep}. The \Xe\ activity concentrations in Flight 2 were
significantly lower than simultaneous ground level air measurements
at Schauinsland.  This may be explained by settling of the heavy xenon
in the atmosphere and is also confirmed by model calculations presented
in \cite{DLR-in-prep}.

\section{Conclusions}

We have reported a new technique successfully combining \Xe\ detection
using miniaturized proportional counters with airplane sampling campaigns.
In particular, we have showed that small air samples of litre scale (STP) are
sufficient to achieve a competitive detection limit in the range of 100
mBq/m$^3$ (STP). This is due to both the ultra-low background rate of the
applied proportional counters which is in the range of few events per day
at 0.5~keV threshold and the highly efficient and contamination-free gas
sample preparation procedure.\\

Our measurements reveal that a part of the Fukushima \Xe\ plume was lifted
to the upper troposphere and to the tropopause. There it was carried by
the fast jet stream to Europe. The earlier arrival time in Germany at
high altitude compared to ground level was unambiguously demonstrated by a
comparison with data from the CTBT Radionuclide Station 33 of the Bundesamt
f\"ur Strahlenschutz on the Schauinsland mountain. We also got evidence for
\Xe\ traces from the Fukushima accident in the upper troposphere during our
second flight on April 14, 2011, although radioactive decay, dilution and
settling of xenon reduced the \Xe\ activity concentration significantly.


\bibliographystyle{elsarticle-num}

\end{document}